\documentclass[11pt]{article}

\usepackage{osidauthor}
\usepackage{graphicx}
\usepackage{young}
\usepackage{amsmath,amssymb,yhmath,makeidx,mathdots,setspace,framed,young,lscape}
\usepackage[enableskew]{youngtab}  
\usepackage[usenames]{color}
\usepackage[OT4]{fontenc}
\usepackage[cp1250]{inputenc}

\newcommand{\trnm}[1]{\mbox{#1}}
\newcommand{\ncd}{\newcommand}

\ncd{\nn}{\nonumber}
\ncd{\ba}{\begin{array}}
\ncd{\ea}{\end{array}}
\ncd{\be}{\begin{equation}}
\ncd{\ee}{\end{equation}}
\ncd{\bea}{\begin{eqnarray}}
\ncd{\eea}{\end{eqnarray}}
\ncd{\ga}{\alpha}
\ncd{\gb}{\beta}
\ncd{\GG}{\Gamma}
\ncd{\gl}{\lambda}
\ncd{\GL}{\Lambda}
\ncd{\grg}{\gk \ximma}
\ncd{\go}{\omega}
\ncd{\GO}{\Omega}
\ncd{\gr}{\rho}
\ncd{\nl}{\nabla}
\ncd{\gre}{\epsilon}
\ncd{\td}{\widetilde}
\ncd{\st}{\longrightarrow}
\ncd{\stk}{\rightarrow}
\ncd{\na}{\longmapsto}
\ncd{\fa}{\forall}
\ncd{\cc}{\circ}
\ncd{\za}{\subset}
\ncd{\wtw}{\Longleftrightarrow}
\ncd{\pcz}{\partial}
\ncd{\gz}{\zeta}
\ncd{\gt}{\theta}
\ncd{\vgt}{\vartheta}
\ncd{\vgr}{\varrho}
\ncd{\gs}{\sigma}
\ncd{\GS}{\Sigma}
\ncd{\cl}{{\cal L}}
\ncd{\gd}{\delta}
\ncd{\GD}{\Delta}
\ncd{\ld}{\trnm{Ldiff}}
\ncd{\cj}{{\cal J}}
\ncd{\ce}{{\cal E}}
\ncd{\cf}{{\cal F}}
\ncd{\tr}{\trnm{Tr}}
\ncd{\bgw}{{\bigwedge}}
\ncd{\im}{\trnm{im}}
\ncd{\szesc}{{\times _6}}
\ncd{\siedem}{{\times _7}}
\ncd{\zlap}{\frac{\pcz ^2}{ \pcz z \pcz z^*}}
\ncd{\ztap}{\frac{\pcz }{ \pcz t}}
\ncd{\pczz}{\frac{\pcz}{\pcz z}}
\ncd{\pczsp}{\frac{\pcz}{\pcz z^*}}
\ncd{\zlapkp}{\frac{\pcz ^2}{ \pcz z_{kp} \pcz {z_{kp}^*}} }
\ncd{\bra}{\langle}
\ncd{\ket}{\rangle}
\ncd{\mnp}{M}
\ncd{\gk}{\overline}
\ncd{\q}{{\mathbb{Q}}}
\ncd{\ccc}{{\mathbb{C}}}
\ncd{\zz}{{\mathbb{Z}}}
\ncd{\qr}{{\q }(\gr)}
\ncd{\zr}{{\bf Z}(\gr)}
\ncd{\cp}{{\bf p}}
\ncd{\cO}{{\cal O}}
\ncd{\ph}{{\cal H}}
\ncd{\din}{\trnm{dim}}
\ncd{\End}{\trnm{End}}
\ncd{\id}{\trnm{id}}
\ncd{\bo}{B\setminus \{ 0\} }
\ncd{\z}{{\mathbb{Z}}}

\title{Galois actions on the eigenproblem\\ of the Heisenberg heptagon}
\author{Jan Milewski \\{\footnotesize\it Institute of Mathematics, Pozna\'n University of Technology,\\
ul. Piotrowo 3A, 60-965 Pozna\'n, Poland \& jsmilew@wp.pl}\\[2ex]
Grzegorz Banaszak \\{\footnotesize\it Department of Mathematical and Computer Science, Adam Mickiewicz University,\\ ul. Umultowska 87, 61-614 Pozna\'n, Poland \& banaszak@amu.edu.pl}\\[2ex]
Tadeusz Lulek\\{\footnotesize\it Faculty of Physics, Adam Mickiewicz University,\\
ul. Umultowska 85, 61-614 Pozna\'n, Poland \& tadlulek@amu.edu.pl}\\[2ex]
Miroslaw Labuz\\{\footnotesize\it Institute of Physics, University of Rzeszow,\\
ul. Rejtana 16a, 35-959 Rzesz\'ow, Poland \& labuz@univ.rzeszow.pl}\\[2ex]
Ryszard Stagraczynski \\{\footnotesize\it Faculty of Mathematics and Applied Physics, Rzeszow University of Technology, \\
ul. Powsta\'nc\'ow Warszawy 6, 35-959 Rzesz\'ow, Poland
}}

\begin{document}


\maketitle

\begin{abstract}
We analyse the exact solution of the eigenproblem for the Heisenberg Hamiltonian of magnetic heptagon, i.e. the ring of $N=7$ nodes, each with spin $1/2$, within the XXX model with nearest neighbour interactions, from the point of view of finite extensions of the field $\mathbb{Q}$ of rationals. We point out, as the main result, that the associated arithmetic structure of these extensions makes natural an introduction of some Galois qubits. They are two-dimensional subspaces of the Hilbert space of the model, which admit a quantum informatic interpretation as elementary memory units for a (hypothetical) computer, based on their distinctive properties with respect to the action of related Galois group for indecomposable factors of the secular determinant.

These Galois qubits are nested on the lattice of subfields which involves several minimal fields for determination of eigenstates (the complex Heisenberg field), spectrum (the real Heisenberg field), and Fourier transforms of magnetic configurations (the cyclotomic field, based on the simple 7th root of unity). The structure of the corresponding lattice of Galois groups is presented in terms of Kummer theory, and its physical interpretation is indicated in terms of appropriate permutations of eigenstates, energies, and density matrices.
\end{abstract}



\section{Introduction}

The matrix elements of the $XXX$ Heisenberg Hamiltonian with the nearest neighbour interactions in the basis of magnetic configurations are integers (in short, the Hamiltonian is arithmetic).
This fact imposes interesting properties of solution of the eigenproblem of the Hamiltonian. Namely, the solution of the eigenproblem is expressible within a {\em finite} extension of the prime field $\q$ of rationals.
It is thus natural to study the Galois symmetry of such an extension. 
In a previous paper \cite{mbll} we have considered some arithmetic aspects of field extensions in the context of pentagon, i.e. a magnetic Heisenberg ring with $N=5$ nodes.
There, it was sufficient to consider the cyclotomic extension by $\exp (2\pi i/5)$, a primitive fifth root of unity. 
The aim of the present paper is to demonstrate a rich theoretic-arithmetic structure of such an extension for the case of magnetic heptagon, that is the ring with $N=7$ spins $1/2$ with nearest neighbour isotropic interaction.
For the case of a single deviation, the eigenproblem is expressible, by means of the Fourier transform, in terms of the cyclotomic field $\q (\go )$, with $\go:=\exp (2 \pi i/7)$. For the heptagon, however, the corresponding algebraic integer $\go$ is not sufficient for a larger number $r$ of spin deviations, and one needs some further extensions by use of Kummer theory \cite{lang}. 

We recall that the $XXX$ model is characterized by the following exact quantum numbers: the quasimomentum $k$, the magnetization $M=7/2-r$, and the total spin $S=7/2-r'$.
The integer $r$ is the total number of overturned spins, whereas $0\leq r'\leq r$ for $0\leq r\leq 3$, has the meaning of those magnons which are coupled into strings \cite{llls}. We demonstrate here that for a fixed triad $(k,r,r')$, $r'>1$, the corresponding secular matrix has exactly the size $2\times 2$, which corresponds in the language of quantum computing \cite{nch} to a specific implementation of a system of qubits within the Hilbert space of the magnetic heptagon.
The Galois group of the global extension turns out to be composed of some Coxeter reflections along the Kummer theory, each reflection acting on a single qubit.

The paper is organized as follows. In Section \ref{prem} we introduce the notation concerning the Heisenberg magnetic ring, adapted from our previous papers \cite{mbll}, \cite{llls}, \cite{mll} - \cite{llm} to the case of heptagon. In Section \ref{gwcf} we introduce the cyclotomic field $\q(\omega)$ which is the first conceptual extension of $\q$, with the physical meaning of quasimomentum and Brillouin zone of heptagon. In Section \ref{eigen} we demonstrate that all two-magnon ($r=2$) and three-magnon ($r=3$) spaces with a fixed quasimomentum $k$ admit natural decompositions in which the highest weight subspaces are two-dimensional. This fact justifies the quantum informatics interpretation of these subspaces, and thus of all relevant spaces for the heptagon, in terms of a collection of qubits. We propose some non-orthogonal bases for these qubits whose components are expressible in the field $\mathbb{Q}(\go)$ with respect to the initial, arithmetic basis of magnetic configurations. Further  we construct projection operators for these qubits in terms of raising and lowering operators for the total spin.
Our construction makes natural a definition of a Galois qubit as one with appropriately restricted field of coefficients.  Section \ref{arit} yields the arithmetic-theoretic analysis of secular equations for our qubits. We prove that the corresponding discriminants are not squares in the field $\q(\rho)$, $\rho=\go+\go^{-1}$, and thus, a square root of each discriminant yields a quadratic extension of $\q(\rho)$. Then in Section \ref{toten} we use the Kummer \cite{lang} theory to demonstrate that the field ${\bf H}_E$ for the heptagon, i.e. the minimal field which encompasses all energies of this magnet (we refer to it as the real Heisenberg number field), has the Abelian Galois group of order $2^6$ over to $\q(\gr)$, and an appropriate wreath product over to $\q$. Further, we describe the corresponding complex Heisenberg number field ${\bf H}_G$ which includes $\q(\go)$, and thus allows to express all density matrices corresponding to exact Bethe Ansatz eigenstates for the heptagon. Section \ref{galqu} demonstrates how to exploit the structure of Galois groups to generate simple transformations between Bethe Ansatz eigenstates.

\section{Preliminaries}
\label{prem}

We briefly recall, using the notation of our previous paper \cite{mbll}, that the eigenproblem of the Heisenberg Hamiltonian for the ring of $N$ nodes, each with the spin $1/2$, is formulated within the Hilbert space 
$\ph \cong \bigotimes ^N \ccc^2$. The space $\ph $ is spanned over the set
\be \td 2 ^{\td N}=\{ f: \td N \st \td 2\} \ee
of all magnetic configurations on the ring $\td N=\{j=1,2,\ldots ,N\}$ ($N=7$ in the present paper), with $\td 2 =\{ 1,2\} \cong \{  1/2,-1/2  \}\cong \{ +,-\} $ denoting the set of $z$-projections of the spin $1/2$; the latter can be readily interpreted in terms of signs $\{\pm \}$, or as the computational basis of the qubit $\ccc^2$, the memory unit of a quantum computer. The unitary structure of the space $\ph $ is defined by assuming the set $ \td 2 ^{\td N}$ as one of its orthonormal bases.

Let $r$ be the number of overturned spins in a magnetic configuration $f\in \td 2 ^{\td 7}$, $Q_r$ be the set of all magnetic configurations with $r$ overturned spins, and
\be \label{lcc} \ph _r=\trnm{lc}_{\ccc} Q_r \ee
be the subspace of $\ph $, spanned on $Q_r$. We recall here that $Q_r$ has the meaning of the classical configuration space of the system $\td r=\{\ga = 1,2,\ldots ,r\}$ of $r$ Bethe pseudoparticles \cite{mll}-\cite{mbl}.

More precisely, each magnetic configuration $f\in Q_r \za \td 2 ^{\td N}$ is presented in the form
\be \label{yb} {\bf j}=(j_1,j_2 ,\ldots ,j_r), 1\leq j_1<j_2<\ldots <j_r \leq N, \ee
where $j_{\ga } \in \td N$ denotes the position of the $\ga$'th overturned spin, counted from the leftmost node $j=1$ to the rightmost $j=N$. This spin deviation is interpreted as the $\ga $'th Bethe pseudoparticle. The inequality (\ref{yb}) imposes the Yang-Baxter structure on the classical configuration space $Q_r$, which acquires the interpretation of a locally hypercubic lattice in $r$ dimensions, with some $F$-dimensional boundaries preventing coincidences of pseudoparticles, $1\leq F<r$. Now, $\ph _r$ becomes the space of all quantum states of the system $\td r$. This space is invariant with respect to the Heisenberg Hamiltonian $\hat H$. So the Hamiltonian is defined by family of 
operators $\hat H _r \in \End (\ph _r)$ which act according to the formula
\be \label{hhr}
\hat H_r |Q, {\bf j}\ket =\sum _{j' \in Q_{r, {\bf j}}} \left(|Q, {\bf j}'\ket-|Q, {\bf j}\ket \right)
,\ee
where $Q_{r, {\bf j}}\za Q_r$ is the set of all nearest neighbours of ${\bf j} \in Q_r$ within the Young-Baxter structure of $Q_r$, and $|Qj\ket$ denotes the quantum state of the system $\td r$, corresponding to its {\it position}, 
specified by ${\bf j}$. Hence, the Heisenberg Hamiltonian $\hat H$ is a direct sum of the family of Hamiltonians $\hat H _r$ 
\be \hat H=\bigoplus _{r=0}^7 \hat H _r. \ee

Observe, that, according to simple combinatorial bijection,
every position ${\bf j}$ in (\ref{yb}) can be treated as an $r$-element subset of $\td{N}.$ Problems related to the classical 
Heisenberg magnetic configuration space with a fixed number of spin deviations were considered in papers
\cite{mll}-\cite{mbl}.
Hence, the vector $|Q,{\bf j}\ket$ can be written in the form
$|\{j_1, \ldots ,j_r\} \ket $. This notation is convenient for the presentation of the step operator $\hat S ^-$ of the total spin of magnet. 
The operator $\hat S ^-$ can be treated in two ways: as an endomorphism of the space $\ph$ or as a family of operators between consecutive subspaces $\ph _r\st \ph _{r+1}$
\be \hat{S}^{-}: \ph _r \st \ph _{r+1}, \; \hat{S}^{-}|\{j_1, \ldots ,j_r\} \ket =\sum_{j\notin \{j_1, \ldots ,j_r\} }|\{j,j_1, \ldots ,j_r\} \ket .\ee
In particular,  ${ \hat{S}^{-}}$ commutes with the Heisenberg Hamiltonian $\hat{H}$, it means that
\be { \hat{S}^{-}}\hat H_r= \hat H _{r+1}{ \hat{S}^{-}}. \ee
This fact allows us to introduce subspaces $\ph _{r,r'}$ of $\ph _r$ in the following way. The subspace $\ph _{r,r}$  (with $r'=r$, called the subspace of the {\it highest weight}) is by definition orthogonal to the image of the operator ${ \hat{S}^{-}}$ in 
$\ph _r$. For $r'<r$ we put
\be \ph _{r,r'}:=  \left( { \hat{S}^{-}}\right)^{r-r'}\ph _{r',r'}. \ee
The subspace $ \ph _{r,r'}$ is the space of quantum states with given number $r$ of deviations and $r'$ of them coupled into strings.


\section{Galois wavelets and algebraic structure of the cyclotomic field}
\label{gwcf}

We would like to put now an emphasis on the fact that Eq. (\ref{hhr}), which is essentially the starting point of the Bethe Ansatz \cite{bethe}, imposes that the matrix elements of the Heisenberg Hamiltonian are integers, and thus the solution
of the corresponding eigenproblem is expressible within a finite extension of the prime field $\q $ of rationals. Thus the algebraically closed field ${\ccc}$ of complex numbers, used in Eq. (\ref{lcc}) along standard rules of quantum mechanics, might be somehow redundant for a complete physical description of the model. Moreover, one can exploit some purely algebraic symmetries of appropriate finite extensions of the prime field $\q$ to learn much about a complete solution of the physical eigenproblem. For example, some Galois automorphisms can define selection rules for transitions between eigenstates.

In this section we apply the Fourier transform in each space $\ph _r$ as the first step in diagonalization of the Hamiltonian $\hat H$ in accordance with papers \cite{mbll}, \cite{llls}, \cite{lb}. To do it, we exploit the orbit structure of each configuration space $Q_r$ with respect to the group $C_N$ - the translational symmetry group of heptagon.
This structure is presented in detail in Table 1, with
\be {\bf t}=(t_1,\ldots ,t_r), \sum _{\ga \in \td r} t_\ga =7,\ee
where
\be
 t_\ga  =\left\{
\ba{lcl}  j_{\ga +1}-j_{\ga}  & , & 1\leq \ga <r\\
                        N+j_1-j_r              & , &  \ga =r
\ea
\right.
\ee
denotes the vector of relative positions of $r$ Bethe pseudoparticles on the heptagon, and characterizes a ${\bf C}_7$-orbit (the characterization is complete when one identifies sequences 
$(t_1,t_2,\ldots ,t_r)\sim (t_2,t_3,\ldots ,t_1)\sim  \ldots \sim  (t_r,t_1,\ldots ,t_{r-1})$ - cf. \cite{mll}-\cite{mbl}.
We use the sequence which is lexically the first, and the integer $1\leq F \leq r$ denotes the number of adjacent islands of Bethe pseudoparticles in the ${\bf C}_7$-orbit ${\bf t}$ \cite{lb}. Table 1. presents only those magnetic configurations for which $r< 7/2$ ("below equator"), and all other configurations emerge from particle-hole symmetry.
Observe, that all orbits are regular (and thus each consists of $7$ configurations), with the exception of $r=0$ (ferromagnetic vacuum) and $r=7$ (anti vacuum).


\begin{table}[htp]
\centering
$
\begin{array}{|c|c|c|c|}
\hline
r& \dim \ph _r&{\bf t}&F\\
\hline
\hline
0&1&\emptyset &0\\
\hline
1&7&7&1\\
\hline
2&21&(1,6)&1\\
 &  &(2,5)&2\\
 &  &(3,4)&2\\
 \hline
3&35&(1,1,5)&1\\
 &  &(1,2,4)&2\\
 &  &(1,3,3)&2\\
 &  &(1,4,2)&2\\ 
 &  &(2,2,3)&3\\
 \hline 
\end{array}
$
\caption{Orbit structure of the classical configuration spaces $Q_r$ for the heptagon, $0\leq r\leq3$. ${\bf t}$ denotes the vector of relative position for the system $\td r$ of Bethe pseudoparticles, and $F$ is the number of islands of adjacent spin deviations.} \label{orbits}
\end{table}

Now we perform the Fourier transform on each such an orbit by means of the formula
\be\label{fowl}  |B,r,{\bf t},k\ket =\sum _{j\in \td{7}} F_{jk} |Q,r,{\bf t},j\ket, k\in B ,\ee
with
\be \label{wspf}  F_{jk} =\go ^{-kj}/\sqrt{7}, j\in \td{7}, k\in B. \ee
Here, $k$ is an integer modulo $7$, with representative taken within the range
\be B:=\{-3\leq k \leq 3\},\ee
and referred to as the {\it qusimomentum}, with the set $B$ recognized as the dual to the translation group ${\bf C}_7$, or the {\it Brillouin zone} for the heptagon. In more detail,  each quasimomentum $k\in B$ defines an irrep
$\GG _k$ of the group ${\bf C}_7$, specified by
\be \GG _k (j) =\go ^{kj}, \quad j\in \td{7} .\ee
The states $|Q,r,{\bf t},j\ket$ constitute the arithmetic basis of ${\bf C } _7$ orbits in $\ph$ along Table 1 (when including states beyond equator), and the states $|B,r,{\bf t},k\ket$ form {\it the normalized basis of wavelets}.
Clearly, each space $\ph_r$ decomposes into ${\bf C}_7$-invariant subspaces $\ph _{r}^{k}$ with a fixed quasimomentum $k$
\be \label{dekompp} \ph _r =\bigoplus_{k\in B}\ph _r ^k.\ee
These subspaces are $1,3$, and $5$-dimensional for $r=1, r=2$ and $r=3$, respectively, and moreover,
the case $r=0$ yields the $1$-dimensional subspace $\ph _{0,0}$, corresponding to the center $k=0$ of the Brillouin zone $B$.
The orthonormal basis (\ref{fowl}) is inconvenient for our purposes since the Fourier coefficient $F_{j,k}$ of Eq. (\ref{wspf}) does not belong to the cyclotomic field $\q(\go )$.
In order to remain within this field, we propose in the following a modified transform
\be |G,r,{\bf t}, k\ket =\sum _{j\in \td{7}} G_{jk} |Q,r,{\bf t},j\ket, \quad k\in B  , \quad G_{jk}= \GG _{-k} (j) , j\in \td{7}\ee
referred hereafter to as {\it  Galois-Fourier} transform with the corresponding basis of {\it Galois wavelets}.
Within this setting, the coefficients $G_{j,k}$ are algebraic integers in $\q (\go)$.

Further, we can make the following decomposition
\be \label{drdeco} \ph _r ^k=\bigoplus _{r'} \ph _{r,r'} ^k , \;  \; \ph _{r,r'} ^k:=\ph _{r,r'} \cap \ph _r^k .\ee
The subspaces $\ph _{r,r'} ^k$ are invariant with respect to the Heisenberg Hamiltonian as the intersection of two invariant subspaces.
The dimensions of these subspaces based on quantum numbers $r,r', k$ are given as follows:
\be \din  \; \ph _{r,r'} ^0=1-r', \din \; \ph _{r,r'} ^l=r' \; \; \trnm{for} \; \; r'=0,1, \trnm{and} \; \; l \in \bo , \ee
\be \label{hqub} \din \; \ph _{r,r'} ^k=2 \; \; \trnm{for} \; \; r'=2,3,  k\in B, \ee
where $r=r' \ldots , 7-r'$.  So the quantum spaces $\ph _{r,r'} ^k$ for $r'=2,3$ are qubits from the quantum computing point of view.

Now let us study some properties of the cyclotomic field $\q (\go )$ from the Galois theory point of view.
The extension $\q (\go ) /\q $ has degree six. 
It means that the field $\q (\go )$ is the linear six-dimensional space over the prime field $\q$, and the set $\{ \go ^k|k\in \bo \}$ of all roots of the minimal polynomial $f_\go$ of $\go $
\be f_\go  (x)=x^6+x^5+\ldots x+1\equiv \prod _{l \in B\setminus \{ 0\} } (x-\go ^l)  \ee
can be used as a basis of this space (root basis), i.e.
\be \q (\go)=\trnm{lc} _\q \{\go ^l : l\in \bo \}, \quad [\q (\go ): \q]=\trnm{deg} f_\go =6 .\ee
The Galois group of the extension $\q (\go ) /\q $ is the group ${\bf C}_6$ isomorphic to the multiplicative group of the ring ${\z }_7$:
\be G(\q (\go )/\q)= {\bf C}_6=\{ \tau _l :l \in {\zz }_7 ^* \} , \quad \tau _l \go ^k= \go ^{lk},\ee
with the multiplication law
\be \tau_{l} \tau _{l'}= \tau_ {l l'} \ee
with $ll'$ understood $\mod 7$.
Observe, that elements $\tau _{-4}$ and $\tau _{-2}$ have order six in ${\bf C}_6$. So, there exist two isomorphisms between the additive group ${\z}_6$ and ${\bf C}_6$
\be \phi _{l}: \z_6 \st {\bf C}_6, \quad l=-4, -2, \quad \phi _l(n)= \tau _{l^n } , \quad n\in \z_6 ,\ee
with $l^n$ understood $\mod 7$.
The isomorphisms are well defined by the Small Fermat Theorem.
The group is the direct product of its two non-trivial subgroups:
\be \label{cdct} {\bf C}_2=\{ \tau _1, \tau_{-1} \} \; \; \trnm{and} \; \;  {\bf C}_3=\{\tau _1, \tau_2, \tau_4\}. \ee
Observe, that ${\bf C}_3$ is the group of squares in ${\bf C}_6$. In accordance with the fundamental theorem of the Galois theory these subgroups determine two subfields of the field $\q (\go)$.
The group ${\bf C}_2$ determines the subfield of invariants of the ${\bf C}_2$. It is the real subfield $\q (\gr )$ of  $\q (\go )$, where $\gr =\go +\go ^{-1}$. The orbits of the action of ${\bf C}_2$ on the root basis are
$ \{\go ,\go ^{-1}\},  \{\go ^{2},\go ^{-2}\},  \{\go ^{4},\go ^{-4}\}$. 
Hence, numbers 
\be \gr _1=\gr= \go +\go ^{-1}, \quad \gr _2= \go ^2 +\go ^{-2}, \quad \gr _4= \go ^4+\go ^{-4} \ee
constitute a basis of the subfield $\q (\gr )$. The subgroup ${\bf C}_3$ acting on basis vector gives two orbits
$ \{\go , \go ^2, \go ^4\}$, $\{\go ^{-1}, \go ^{-2}, \go ^{-4}\} $.
So 
\be \eta _1=\go + \go ^2 +\go ^4, \quad \eta _{-1}=\go ^{-1}+ \go ^{-2}+ \go ^{-4} \ee
is the basis of ${\bf C}_3$ invariants.
Clearly, the subfield of ${\bf C}_3$-invariant is $\q (\eta)$, with
\be \eta:=\eta_1-\eta_{-1}=i\sqrt{7} .\ee

\begin{figure}[h]
\begin{center}
\includegraphics*[width=3.5cm]{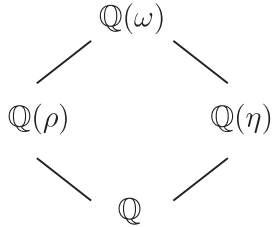}
\end{center}
\caption{The lattice of subfields for heptagon.}
\label{fig:sublat}
\end{figure}

The lattice of subfields of the cyclotomic fields $\q (\go)$ is given on Fig. \ref{fig:sublat}.\\

\newpage

\section{The eigenproblem and Galois qubits}
\label{eigen}

In accordance to the decomposition (\ref{dekompp}), the Hamiltonian $\hat{H} _r$ can be presented as a direct sum
\be
\hat{H}_r = \bigoplus_{r,k} \hat{H}_r ^k  .
\ee
Correspondingly, the secular matrices $H_{r,k}$ in the basis of wavelets read
\be \label{ham0} H_0 ^0=[0], \ee
\be\label{ham1} H_1 ^k=[-2+\gk \xi +\xi] \, ,
  \ee
\be 
\label{ham2}
H_2 ^k=
\left[
\ba{ccc}
-2  &1+\xi  & 0 \\
1+\gk \xi  &-4 &  1+\xi   \\ 
 0 &  1+\gk \xi   &-4+ \xi ^3+\gk \xi ^3   
\ea
\right] ,
\ee
\be
\label{ham3}
H_3 ^k=
\left[
\ba{ccccc}
-2  & 1  &  0  &  \gk \xi     &  0           \\
 1  &-4  &  1  &  \gk \xi ^2  &  \xi   \\
 0  & 1  & -4  &   1      &  1+\xi ^3 \\
\xi & \xi ^2 & 1 &-4  & \xi ^2 \\
0 & \gk \xi  & 1+\gk \xi ^3   & \gk \xi ^2 & -6 +\xi ^2 +\gk \xi ^2 
\ea
\right], 
\ee
where 
\be \xi:= \go ^k , \quad k\in B.\ee
The rows and columns of matrices in (\ref{ham1}-\ref{ham3}) are labeled by vectors ${\bf t}$ of relative positions 
in accordance with Table $1$.

In order to diagonalize the matrices (\ref{ham2}, \ref{ham3}) we start from the following statement. If $|\Psi \ket \in \ph _r ^k$ is  an eigenstate of the Hamiltonian $\hat{H}$, then $\hat{S} ^{-} |\Psi \ket$ is  an eigenstate of $\hat{H}$ with the same eigenenergy.

Let us define operators
\be S_{r,\GD r}^k:=(S^{-})^{\GD r} P_r^k \ee
by means of the $\GD r$-th power of the creation operator $S^{-}$ of Bethe pseudoparticle and the projection operator $\hat P _r ^k$ on the subspace $\ph _r^k$.
The operator  $S_{r,\GD r}^k$ is an endomorphism of $\ph $
or a homomorphism from  $\ph _r^k$ to $\ph _{r+\GD r}^k$ which increases
the number of deviations and preserves  quasimomenta.

The matrices of these operators for $r=1,2 ,\GD r=1$ and $r=1, \GD r=2$ 
in the Galois wavelets basis have the following forms
\be \label{spp} 
 S_{1,1}^k= 
\left[\ba{c}
1+\gk \xi  \\
1+\gk \xi   ^2\\
1+\gk \xi ^3 
\ea 
\right], \quad
 S_{2,1}^k=
 \left[\ba{ccc}
1+\gk \xi  & 1 & 0 \\
1    & \gk \xi & 1  \\
1 & 0 &\xi ^3+\gk \xi \\
1 & \xi ^2 & \xi ^2 \\
0 & 1 +\gk \xi ^2 & \xi ^3
\ea 
\right]\ee
and
\be \label{spo}
S_{1,2}^k=S_{2,1}^kS_{1,1}^k .\ee
 

Hence,
\be \label{jmag} { \hat{S}^{-}} |G,1,k,7\ket =\sum_{t=1}^3 (1+\xi ^{-kt})|G,2,k,t,7-t\ket =
 \left[\ba{c}
1+\gk \xi  \\
1+\gk \xi   ^2\\
1+\gk \xi ^3 
\ea 
\right]
\in {\cal H}_{2,r'}^k ,\ee
is the zero-magnon ($r'=0$) for $k=0$ and one-magnon ($r'=1$) for $k\neq 0$ eigenstate in the space of two spin deviations, with the eigenvalue 
\be E_{r'}^k=-2+\xi +\gk \xi .\ee
The highest weight two-magnon space $\ph _{2,2}^k$ is two dimensional,  as a subspace of $\ph _2 ^k$, orthogonal to the vector ${ \hat{S}^{-}} |G,1,k,7\ket$.
We refer hereafter to this space as the two-magnon qubit (for given $k \in B$).

Now let us give the solution of the eigenproblem of (\ref{ham2}) for $k=0$. As the zero-magnon state in $\ph _{2,0}^0$ we put
\be \label{dwazero} 
v_{2,0}^0=\frac{1}{2} { \hat{S}^{-}} |G,1,k,7\ket =\left[ \ba{c} 1 \\1 \\1 \ea \right]. \ee
The following vectors  
\be \label{dwadwa}
 v_{2,2;1}^0=\left[ \ba{c} -1 \\0 \\1 \ea \right] \; \;  \trnm{and} \; \;  v_{2,2;-1}^0=\left[ \ba{c} 1 \\-2 \\1 \ea \right]  \ee
which are orthogonal to $v_{2,0}^0$ are eigenstates of (\ref{ham2}) of the highest weight with eigenenergies
\be  E_{2;1}^0=-2  \quad  \trnm{and} \quad  E_{2;-1}^0=-6.\ee

Now let us deal with the eigenproblem of (\ref{ham2}) for $k\neq 0$.
Vectors
\be\label{dwanw} v_{2,1}^k:=
 \left[\ba{c}
1+\xi ^2\\
-(1+\xi )\\
0
\ea 
\right], \; 
 v_{2,2}^k:=\frac{1}{1+\xi }
 \left[\ba{c}
1+\xi ^3\\
0\\
-(1+\xi )
\ea 
\right] =
 \left[\ba{c}
1+\xi +\xi ^2\\
0\\
-1
\ea 
\right] 
\ee
constitute a basis of the highest weight two-magnon space as orthogonal to the subspace generated by the one-magnon state (\ref{jmag}). The construction of these vectors is based on the first step of the Jacobi's method. Observe, that
vectors (\ref{dwanw}) are not orthogonal. 
The two-magnon Hamiltonian in the basis (\ref{dwadwa}) 
has the form
\be \label{ham22}
H_{2,2}^k=
\left[
\ba{cc}
-\mu &  2-\mu\\
2+\mu & \mu_4   \ea
\right]
-4I_2,
\ee
where $\mu =\xi +\gk \xi,$ $\mu_l=\xi^l+\xi^{-l}.$ Observe, that $\mu_{7-k} = \mu_{-k}=\mu_{k}$.

Let $k\neq 0$. Then, the minimal polynomial for $\mu $ has the form
$w(x)=x^3+x^2-2x -1$, and so we get

\be \label{ro34} \mu^3=1+2\mu -\mu ^2 \, , \quad \mu^4=-1-\mu +3\mu ^2.\ee
The  characteristic polynomial of the operator $\hat H_{2,2}^k$ is
\be \label{wiel2} f_{2,2}^k(x)=(x+4)^2-(1-2\mu -\mu ^2)(x+4)+(-3+\mu +\mu ^2) . \ee
with discriminant of the characteristic polynomial
\be \GD _2^k=16-\mu-3\mu^2  \ee
so that the two-magnon energies read 
\be \label{e2}
 E_{2;\pm 1}^k=\frac{1}{2} (1-2\mu-\mu ^2 \pm \sqrt{\GD _2 ^k})-4  .\ee

Now let us consider the three-magnon problem. At first let us observe, that the image of the operator $\hat S ^{-}$ in the space $\ph _3^ k$ is generated by vectors 
$f_1:={ \hat{S}^{-}} |G,2,k,(1,6)\ket$, $f_2:={ \hat{S}^{-}} |G,2,k,(2,4)\ket$, $f_3:={ \hat{S}^{-}} |G,2,k,(3,4)\ket $ 
which  in the Galois wavelets basis have the form
\be 
\label{fff}
f_1=\left[\ba{c} \gk \xi+1 \\ 1 \\ 1 \\ 1 \\ 0  \ea \right],
f_2=\left[\ba{c} 1 \\ \gk \xi \\ 0 \\ \xi ^2 \\ 1+\gk \xi ^2  \ea \right],
f_3=\left[\ba{c} 0 \\ 1 \\\xi^3+\gk \xi \\ \xi ^2 \\  \xi ^3  \ea \right].
\ee
For $k=0$, using the Gauss elimination method for the study of orthogonality to vectors (\ref{fff}) we get the general form of the highest weight three-magnon vectors 
\be v=
\left[
\ba{c}
2t\\ -3t+s \\ 2t \\-3t-s \\2t 
\ea
\right] \, , \quad t,s \in {\bf C}.
\ee
which results in eigenenergy $E_3^0=-5$ (double degeneration). So 

\begin{equation}
\label{w30}
\begin{array}{l}
v_{3,0}^0=\left[\ba{c} 1 \\1 \\ 1 \\1 \\ 1 \ea \right], \quad
v_{3,2;1}^0=\left[\ba{c} -2 \\0 \\ 1 \\ 0 \\ 1 \ea \right], \quad
v_{3,2;2}^0=\left[\ba{c} 0 \\ 0 \\ -1 \\ 0 \\ 1 \ea \right],\\[+10ex]
v_{3,3;1}^0=\left[\ba{c} 2 \\3 \\ -2 \\ 3 \\ 2 \ea \right], \quad
v_{3,3;2}^0=\left[\ba{c} 0 \\ 1 \\ 0 \\ -1 \\ 0 \ea \right]
\end{array}
\end{equation}
are eigenstates, where those of lower weight are obtained by means of the action $\hat{S}^-$ on (\ref{dwazero}, \ref{dwadwa}).


Now let us deal with the three-magnon eigenproblem for $k\neq 0$.  The three-magnon space of the highest weight 
$\ph _{3,3}$ is the kernel of the operator
$\hat S^{+} :={\hat S^{-}\;}^{\dagger}$. Hence, it is generated by vectors of the form
\be \label{trkon}  (| j_1 \ket- |j_2 \ket ) \odot (|j_3 \ket- |j_4 \ket )\odot  (|j_5 \ket- |j_6 \ket ), \ee
where $j_1,\dots ,j_6$ are pairwise different (it means $\trnm{card} \{j_1,\dots ,j_6\}=6$ . The multiplication $\odot$ is defined at first for $A_1, \ldots , A_l \za\{1, \ldots , \td  N\}$ 
pairwise disjoint: $|A_1\ket \odot \ldots \odot |A_l \ket =|A_1 \cup \ldots \cup A_l \ket $. Further, we extend this multiplication to linear combinations over ${\bf C}$ with pairwise disjoint supports
in the natural way (of course $|j\ket =| \{ j \} \ket$).

Configuration of the form (\ref{trkon}) can be uniquely labeled by the Young tableau

\be
\begin{array}{c}
\begin{Young}
\scriptsize { $j_1 $}  & {\scriptsize $j_3$ } &
{\scriptsize $j_{5}  $}
&{\scriptsize $j_{7 }$}  \cr
{ \scriptsize $j_2$}      & {\scriptsize $j_4$ } & {\scriptsize $j_6$}\;  \cr
\end{Young}\\
\end{array}
\ee
 and the set of all such standard Young tableaux determines a basis in the three-magnon space.

For $k\in B\setminus \{0\}$ we put vectors
\bea \label{trzynw1}
v_{3,1}^k&=&\xi ^3\sum _j \xi ^{-j}(|2+j \ket- |1+j \ket )\odot  (|4+j \ket- |3+j \ket )\odot  (|6+j \ket- |5+j \ket )\nonumber\\
&=&  
\left[
\ba{c}
0 \\-(1+\gk \xi ^2)\\
-(\xi ^2 -\gk \xi ^2)\\
1+\xi ^2\\
\gk \xi -\gk \xi ^2
\ea
\right] 
\eea
and
\bea \label{trzynw2}
v_{3,2}^k&=&(1+\xi)\sum _j \xi ^{-j}(| 1+j\ket- |4 +j\ket )\odot  (|2+j \ket- | 5+j\ket )\nonumber \\
&&\odot\,  (|3 +j\ket- |6 +j\ket)= 
\left[
\ba{c}
\xi-\xi ^5\\
\xi ^5+ \xi ^6\\
0 \\
-(\xi +\xi ^2)\\
\xi ^3-\xi
\ea
\right] 
\eea
which constitute a basis in the three-magnon qubit $\ph _{3,3}^k$.

In this basis the three-magnon Hamiltonian is expressed by the following matrix 
\be \label{ham33}
\hat H_{3,3}^k=
\left[
\ba{cc}
-3-\mu _4  &    -1+\mu_2-2\mu_4 \\
1+\mu_2    &  1+\mu_2 \ea
\right]
-4I_2.
\ee

The characteristic polynomial of the Hamiltonian (\ref{ham33}) has the form
\be \label{wiel3} f_{3,3}^k (x)= (x+4)^2-(-5+\mu +2\mu  ^2)(x+4)+(\mu-2\mu ^2) \ee
and its discriminant is
\be \GD_3^k=25-10\mu-3\mu ^2 .\ee
Three-magnon energy of the highest weight for $k\neq 0$ is given by the following equality  
\be \label{e3}
 E_{3;\pm 1}^k=\frac{1}{2} (-5+\mu +2\mu  ^2 \pm \sqrt{\GD _3 ^k})-4  .\ee

In the case $k=0,$ beyond eigen-energies, we presented also eigenvectors in a simple form. 
For $k\neq 0$ we presented eigenenergies (\ref{e2}, \ref{e3}) for the states of weight $r'$ larger than  $1$ 
(for the weight $r'=1$ the energy is given by the formula (\ref{ham1})).
We will find the density matrix corresponding to an eigenvector. 
For this reason we write down the projectors  $P_{r,r'}^k$  
on the subspaces with given quantum numbers $r,r',k$ for $k\neq 0$.

At first let us observe, that the traces of products of the matrices (\ref{spo}, \ref{spp}) and their Hermitian conjugations for $k\neq 0$ are
\be \tr \,  S_{1,1}^k S_{1,1}^{k\; \dagger} =5, \quad  \tr \,  S_{2,1}^k S_{2,1}^{k\; \dagger} =14, \quad \tr \,  S_{1,2}^k S_{1,2}^{k\; \dagger} =40. \ee
Hence, projection operators in the two-deviation space with given wave vector $k\neq 0$ on one-magnon and two-magnon space are:
\be \label{prp}
P_{2,1}^k=\frac{1}{5} S_{1,1}^kS_{1,1}^{k \; \dagger}. \ee
The projection operator $P_{2,1}^k$ is the density matrix because the space $\ph _{2,1}^k$ is one-dimensional
\be  P_{2,2}^k=P_2^k-P_{2,1}^k .\ee
The projection operators in the three-deviation space with given wave vector $k\neq 0$ onto one-magnon, two-magnon and three-magnon space are given by the following formulas
\be P_{3,1}^k=\frac{1}{40} S_{1,2}^kS_{1,2}^{k \; \dagger} , \ee
\be P_{3,2}^k=\frac{1}{3} S_{2,1}^kS_{2,1}^{k \; \dagger}-\frac{1}{15} S_{1,2}^kS_{1,2}^{k \; \dagger}  ,\ee
\be   \label{pro}
 P_{3,3}^k=P_{3,k}  - \frac{1}{3} S_{2,1}^kS_{2,1}^{k \; \dagger}+\frac{1}{24} S_{1,2}^kS_{1,2}^{k \; \dagger} .\ee 
For $r'=1$ the density matrices are given by
\be \vgr _{r,r'}^k =P_{r,r'}^k \ee
as the spaces $\ph _{r,r'} ^k$ are one-dimensional. 

For $r'=2,3$ the density matrices $\gr _{r,r',1}^k$ and $\gr _{r,r',2}^k$ satisfy the following system of linear equations:
\be \vgr_{r,r',-1}^k+\vgr_{r,r,1}^k=P_{r,r'}^k \, , \quad   E_{r',-1}^k \vgr_{r,r',-1}^k +E_{r',1}^k \vgr_{r,r',1}^k  =H_{r,r'}^k , \ee
where 
$ E_{r',\pm1}^k$ are energies of two- and three-magnon eigenstates given by (\ref{e2}, \ref{e3})  and
\be H_{r,r'}^k= H_{r}^kP_{r,r'}^k . \ee 
The solution of the system is 
\be \vgr_{r,r',\nu}^k=\nu \frac{H_{r,r'}^k -E_{r',-\nu}^kP_{r,r}^k}{\sqrt{\GD_r ^k}}  \; , \quad  k= \pm 1, \pm 2,\pm 4  , \ee
where $\nu=\pm 1$ is a digit for the qubit $\ph _{r,r'}^k$, and the eigenenergies are invariant with respect to 
the reflection of the number $k$
$E_{r', \nu}^{-k}=E_{r', \nu}^k$, $\GD_{r'}^{-k}=\GD_{r'}^k$.
In the basis of Galois wavelets Hamiltonians $H_r^k$ are given by (\ref{ham2}, \ref{ham3}) and projection operators
$P_{r,r'}^k$ by (\ref{prp}-\ref{pro}).

For $k=0$ the density matrices and projectors are given by eigenvectors (\ref{w30}) as follows:
\be \vgr_{r,r',\nu} ^0= \frac{| v_{r,r',\nu }^0 \ket \bra  v_{r,r',\nu }^0  |}{ \bra  v_{r,r',\nu }^0  
| v_{r,r',\nu }^0 \ket }, \;\;\; P_{r,r'}^0= \sum _{\nu} \vgr _{r,r',\nu} ^0, \ee
where $\nu $ is the empty index for $r'=0$ and runs through $\pm 1$ for $r'=2,3.$

It is worth to observe, that components of vectors (\ref{dwanw}) and (\ref{trzynw1}, \ref{trzynw2})  in the Galois wavelets (hence, in the arithmetic basis) belong to the field $\q (\go)$. It allows us to propose the following definition. 
\medskip

\noindent
{\bf Definition 1.} {\it Let $K$ be an algebraic extension of $\q$ and $r'=2,3$. Then the linear space: 
\be  {\cal Q} _{r,r'}^k(K):= \{ v \in \ph _{r,r'}^k: \bra v |{\bf j}\ket \in K\, ,{\bf j} \in Q_r\} \ee
over the field $K$ will be called the {\bf Galois qubit} if $\din  _K {\cal Q} _{r,r'}^k(K)=2$.}
\medskip

\noindent
For example $ {\cal Q} _{r,r'}^k(\q)=\{0\}$ for $k\neq 0$ and $\din  _\q {\cal Q} _{r,r'}^0(\q)=2$. Hence, $ {\cal Q} _{r,r'}^k(\q)$ is the Galois qubit only for $k=0$. For $k\neq 0$ $ {\cal Q} _{r,r'}^k(K)$
is the Galois qubit over $K$ if the field $K$ is an extension of the cyclotomic field $\q(\go )$.
In particular,  ${\cal Q} _{rr'k}(\q (\go ))$ will be referred to as {\bf Galois-Fourier qubit}.
Elements of the matrices (\ref{ham22}, \ref{ham33}) belong to the field 
$\q (\mu )\za \q (\go)$ so the two- and three-magnon 
Hamiltonians are endomorphisms of qubits $\ph  _{rr'}^k$ and the Galois qubit 
${\cal Q} _{rr'}^k(\q (\go))$ as well.



\section{Arithmetic properties of two and three-magnons discriminant}
\label{arit}

The secular equations for two and three magnons with given $k$ are quadratic.
Discriminants $\Delta_{r'}^k$ of these equations are
\be \label{twodis}
\GD_{2}^1=16-\gr -3\gr ^2 \; \; , \quad 
\GD_{2}^2=9+3\gr +2\gr ^2 \; \; , \quad
\GD_{2}^4=9-2\gr +\gr ^2, \ee
\be \label{thrdis}
\GD_{3}^1=25-10\gr -3\gr ^2 
 \; \; , \quad 
\GD_{3}^2=36+3\gr -7\gr ^2 
 \; \; , \quad 
\GD_{3}^4=9+7\gr +10 \gr ^2. \ee
Each solution of an eigenproblem is expressible in terms of a square root of appropriate discriminant. 
It is natural to ask if square roots of these discriminants belong to the field $\q  (\gr)$. The following theorem gives the negative answer.
\medskip

\noindent
{\bf Theorem 1.} {\it The discriminants $\GD _{r'}^k$ for $r'=2,3$ and $k=1,2,4$ are not squares in $\q(\gr)$.}
\medskip

\noindent
Proof.
Observe, that
\be\label{fdg}\tau ^l \GD _r^k=\GD_r ^{2^l k}\ee
where $\tau:=\tau_2$ is the generator of the Galois group $G(\q (\rho )/\q)$ (cf. Eq. \ref{cdct}).
It is convenient to use the following notation
\be a_1:=a \; , \quad  a_2=\tau (a) \; , \quad a_4=\tau ^2 (a)\ee
for any $a\in \q(\gr)$. The trace and the norm of an element $a\in K$ for the extension
$K/\q$ are given by means of this notation as follows:
\be \tr_{K/\q} (a)=a_1+a_2+a_4\; , \quad  N_{K/\q} (a)=a_1a_2a_4 ,\ee
where $K:=\q(\gr)$.

As the first step of the proof of this theorem we formulate the technical

{\bf Fact.} {\it For any $x,y\in \q, a\in \q(\gr)$ the following equalities hold}
\be \label{wzskrmn} N_{K/\q} (x+a)=x^3+\tr _{K/\q}\, a \, x^2+\frac{1}{2}[(\tr_{K/\q}\, a)^2-\tr_{K/\q}\, (a^2)]x+N_{K/\q} (a) ,\ee
\be \label{swzskm} N_{K/\q} (x+y\gr)=x^3-x^2y-2xy^2+y^3. \ee

For the proof of (\ref{wzskrmn}) one can use the identity
\be  N_{K/\q} (x+a)=(x+a_1)(x+a_2)(x+a_4) .\ee
The equation (\ref{swzskm}) is a particular case of (\ref{wzskrmn}), because
\be \tr _{K/\q} (\gr )=-1, \quad  \tr_{K/\q}(\gr ^2)=-3, \quad  N_{K/\q}(\gr )=1.\ee

{\bf Lemma 1.} \label{lem1} {\it The norm of the two-magnon discriminants (\ref{twodis}) is a prime number $p_2:=1289$.}

Proof.
The formula (\ref{fdg}) implies that the norms of all these discriminants are equal to each other.
Observe, that   
\be \GD_2^4=8+a  , \ee
where
\be a=(\gr-1)^2=\gr_2-2\gr_1+3 ,\ee
hence,
\be \tr _{K/\q} \, a=10 .\ee
The trace of the square of $a$ we calculate in a similar way
\be a^ 2=(\gr-1)^4=13\gr_2-13\gr_1+22, \ee
so
\be \label{tr66} \tr _{K/\q} (a^2)=66. \ee
Substituting (\ref{tr66}) to (\ref{wzskrmn}) we get
\be\label{norma2}  N_{K/\q} (\GD_{2,3})= 1289 .\ee

{\bf Lemma 2.} {\it The norm of the three-magnon discriminants (\ref{thrdis}) is a composite number 
$n_3:=7\cdot 13 \cdot 83$.}

Observe, that
\be \GD_3^1=(5-3\gr)(5+\gr) .\ee
Then, using (\ref{swzskm}) we get
\be \label{n3}  N_{K/\q} (5-3\gr)=83 \; , \quad N_{K/\q} (5+\gr)=91=7\cdot 13 .\ee

Lemmas 1 and 2 show that the norms of the discriminants are not squares in $\q$, hence, these discriminants are not squares in $K$. This completes the proof of the theorem.

The eigenenergies for given $r'$ and $k$ belong to the 
number field:
\be {\bf H}_{r', E}^k := \q(\gr , \sqrt{\GD_{r'}^k}) .\ee

\noindent
{\bf Definition 2.}
{\it The number field ${\bf H}_{r', E}^k$ will be called the real {\bf Heisenberg number field} for $r'$ and $k$. }
\medskip

\noindent
${\bf H}_{r', E}^k$ is the minimal field which contains the energy of the Hamiltonian ${\hat H}_{r,r'}^k$. Hence, 
$[{\bf H}_{r', E}^k:\q(\gr )]=2$ for $k\in B\setminus \{0\}$, $r'=2,3$ and ${\bf H}_{r', \, E}^0 =\q$.

{\bf Lemma 3.} {\it The two-magnon discriminants are primes in the ring $\cO_K$ of integers in the field $K$. The decomposition of the three-magnon discriminants into primes in the ring $\cO _K$ are
given by the following formulas
\be \label{m31} \GD_{3}^1=(5-3\gr)\tau(3+\gr)\tau(2-\gr) ,\ee
\be \GD_{3}^2=\tau(5-3\gr)\tau^2(3+\gr)\tau^2(2-\gr) ,\ee
\be \label{m33} \GD_{3}^3=\tau^2(5-3\gr)(3+\gr)(2-\gr) .\ee }

Proof. The first part of Lemma 3 is a corollary from  Lemma 1. The equalities (\ref{n3}) show that $5-3\gr$ is a prime, and $5+\gr$ is not prime in  $\cO _K$.  
Now observe, that the following equality
\be \tau ^2(5+\gr)=(3+\gr)(2-\gr) \ee
holds and the norms of these factors are prime numbers:
\be N_{K/\q} (3+\gr)=13 \; , \quad N_{K/\q}(2-\gr)=7. \ee
This finishes the proof of the lemma 3.


\section{The real and the complex Heisenberg number fields}
\label{toten}

\noindent
{\bf Definition 3.}
{\it Let $r'$-magnon {\bf real Heisenberg number field} be the following
composite of fields:  
\be {\bf H}_{r',E} := {\bf H}_{r',E}^1{\bf H}_{r',E}^2{\bf H}_{r',E}^4\ee
and let {\bf total real Heisenberg number field} be defined as another composite of fields:
\be {\bf H}_E := {\bf H}_{2,E}{\bf H}_{3,E}. \ee } 
\noindent
The fields ${\bf H}_{r',E}^k$, ${\bf H}_{r',E}$ and   ${\bf H}_E$  are minimal ones which contain eigenenergies $ E_{r',\pm1}^k$,
all eigenenergies with given $r'$, and all  eigenenergies of the Heisenberg Hamiltonian $\hat H$, respectively.

Now let us define the complex Heisenberg number fields as the minimal  extensions of real Heisenberg fields and the cyclotomic fields. 
So ${\bf H}_{r',E}^k$ is the minimal field for diagonalization of the Heisenberg Hamiltonian ${\hat H}_{r,r'}^k$ in the qubit $\ph _{r,r'}^k$, 
where $r'\in 2,3 $, $k \in B\setminus \{ 0 \}$. 
These fields are given by the following formula
\be {\bf H}_{\GL ,G}={\bf H}_{\GL ,E}(\eta) , \ee
where $\eta=i\sqrt{7}$ for $\GL=r',k$ ( $r'=2,3, k\in B\setminus \{0\}$), or $\GL =r'$ ($r'=2,3$) or $\GL$ being empty index.
Hence,
\be G( {\bf H }_{\GL ,G}/\q (\go ))=G({\bf H}_{\GL ,E}/\q (\gr)).\ee \\
\begin{figure}[h]
\begin{center}
\includegraphics*[width=8cm]{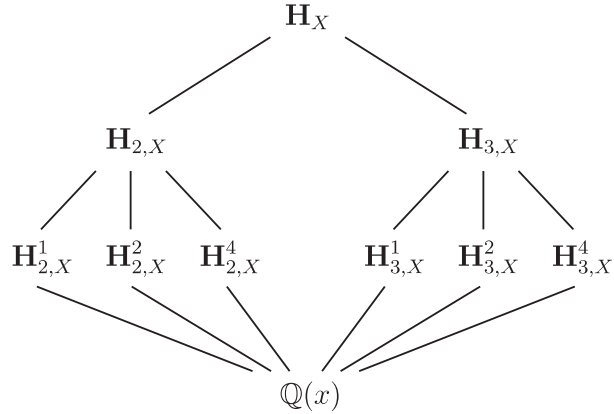}
\end{center}
\caption{The lattice of Heisenberg subfields for heptagon: $X=E$, $x=\rho$ (real case), $X=G$, $x=\omega$ (complex case).}\label{fig:sublat2}
\end{figure}

Now we deal with Galois properties of the extensions ${\bf H}_{r,E}/ \q$, $r=2,3$ and ${\bf H}_E/ \q$.
At first we study the extension ${\bf H}_{r',E}/ \q(\gr )$, ${\bf H}_E/ \q(\gr )$, 
and further, we compose the extension with $\q (\gr)/ \q$.

To investigate   ${\bf H}_{r',E}/ \q(\gr )$ and ${\bf H}_E/ \q (\gr)$ we use the Kummer theory.

In order to make the investigation explicit, let us recall some general facts from algebraic 
number, Galois and, in particular, Kummer theory. 
Let $K$ be a field, $\GD_1, \ldots ,\GD _l \in K$ and $\GD_1, \ldots ,\GD _l \notin K^2$.
Consider the extension $K_l:=K(\sqrt{\GD_1}, \ldots ,\sqrt{\GD _l})$ and multiplicative 
subgroup $\td \GL$ of $K^\times$ generated by $\GD_1, \ldots ,\GD _l $ and ${K^\times}^2$
\be \td \GL=\bra \{ \GD_1, \ldots ,\GD _l\} \cup {K^\times}^2 \ket .\ee
The subgroup is a $\z$-module as an abelian group. 
So
\be \GL=\td \GL / K^{\times \, 2}\ee
is a $\z_2$-module. 
Now, let us assume $\GD_1, \dots, \GD_k \in \cO_K.$ Let $\cp_1, \dots, {\cp}_k$
be a set of prime ideals of ${\cO}_K$. Assume that this set of prime ideals of $\cO _K$
is such that 
\be \label{warkum} \cp _i  | (\GD_i )\; \;  \trnm{and} \; \; \cp _j \not | (\GD _i) \; \trnm{for} \; i\neq j. \ee
Hence, using uniqueness of decomposition into prime ideals in ${\cO}_K$
one gets that $\GL$ is a free $\z/2$ module with basis 
$\GD_1 {K^\times}^2, \dots, \GD_l {K^\times}^2$.
By Kummer theory \cite{lang} ($\S$ 8, Theorem 13) we have bilinear perfect pairing:
\be \bra , \ket :
G(K_l/K) \times \GL \st \z_2\ee 
given by 
\be \bra g,\gl \ket=g\ga/\ga \; , \; \trnm{ where} \;  \ga \in K_l, \ga ^2=\gl  \ee
and 
\be \label{rkt} G(K_l / K) \quad \cong \quad \GL
\quad \cong \quad (\z _2)^l.\ee
In particular it shows that $[K_l \,\, :\,\, K] \,\, = \,\, 2^l $ and \\
$\GD_i \notin K( \GD_{1}, \dots, \GD_{i-1}, \GD_{i+1}, \dots, \GD_{l})$ for every $i = 1, 2, \dots, l.$

The Galois group of the extension acts in the following way:
\be 
\begin{array}{l}
(\gre _1 , \ldots ,\gre _l)(\sqrt{\GD_i})=\gre _i \sqrt{\GD _i} \; \; , \; \;
i=1,\ldots l \; \;, \; \;  \gre_i\in \z_2=\{\pm 1\}.
\end{array}
\ee
Hence, the bilinear pairing is given by
\be \bra (\gre _1 , \ldots ,\gre _l),{\GD_1}^{\ga _1}\cdot \ldots \cdot {\GD_l}^{\ga _l}\ket=
{\gre _1}^{\ga _1}\cdot \ldots \cdot {\gre _l}^{\ga _l}.\ee
 
\noindent Applying these facts to fields extensions
\be \label{fext} {\bf H}_E/\q (\gr) \; , \; \; {\bf H}_{2,E}/\q(\gr ) \;  ,\; \;  {\bf H}_{3,E}/\q(\gr ) \; , \ee
we get the following theorem:

{\bf Theorem 3.} {\it The Galois groups of the extensions (\ref{fext}) are as follows:
\be \label{gal23} G({\bf H}_{2,E}/\q(\gr ))=G({\bf H}_{3,E}/\q(\gr ))={{\bf C}_2}^3  \ee
and
\be \label{galtote} G({\bf H}_E/\q(\gr ))= {{\bf C}_2}^3 \times {{\bf C}_2}^3.\ee}


Proof. In order to prove it we put $K=\q(\gr)$ and set up the following ideals:  
${\bf p}_2^k=(\GD_2^k)$ and ${\bf p}_3^k=(\tau ^k(3+\gr))$ 
(see (\ref{m31})--(\ref{m33})) for $k=1,2,4$  in ${\cO}_K$.
The Galois group $G(\q (\gr )/\q)$ permutes cyclically elements of the 
form $\GD_2^k$ and $\tau^k(3+\gr)$ as well. 
The norms of $\GD _2^1$ and $3+\gr$ are prime numbers mutually different 
and different from  $p=7$.  Moreover, 
\be N_{K/\q} (\GD _2^1) \equiv 1 \mod 7, \;\;\;  N_{K/\q}(3+\gr)\equiv -1 \mod 7. \ee
So the prime ideals $(\GD _2^1)$ and $(3+\gr)$ are split completely in the 
field extension $\q(\gr )/\q$. It follows that \be {\bf p}_{r'}^k | 
\GD _{r'}^k \; \; \trnm{and} \; \; {\bf p}_{r'}^k \not| \, 
\GD _{r'_1}^{k_1} \; \trnm{for}  \; r' _1 \neq   r'  \; \trnm{or} \; k_1 \neq k .\ee
Using the result (\ref{rkt}) of the Kummer theory we get (\ref{gal23}, \ref{galtote}).



Let us remind the definition of the wreath product. Let $G$ and $H$ be groups and $S$ be a set with a representation
$\phi\, :\, H \rightarrow Sym (S).$ Let $G^S := \prod_{s \in S} \, G.$  The wreath product $G \wr_{\phi} H := G^S \rtimes H$ with the composition law:
$$((g_s), h) ((g^{\prime}_{s}), h^{\prime} )= ((g_{\phi (h^{\prime})(s)})(g_{s}^{\prime}), h h^{\prime})$$
for all $(g_s),  (g^{\prime}_{s}) \in G^S$ and all $h, h^{\prime} \in H.$
\medskip

The Galois group of the extensions of total real Heisenberg field over $\q$
is (as a set) the Cartesian product
\be {{\bf C}_2}^3 \times {\bf C}_3. \ee
The action of an element $(\gre_1 ,\gre _2 ,\gre _4; \tau_l)$ on $\gr _k$ and the  square of a discriminant is given as follows

\be (\gre_1 ,\gre _2 ,\gre _4; \tau_l) \gr _k=\gr _{l k} \; \; , \; \; (\gre_1 ,\gre _2 ,\gre _4; \tau_l)
 \sqrt{\GD _{r'}^k}=\gre _l\sqrt{\GD _{r'}^{lk}}. \ee

\noindent
So the Galois group is the wreath products with the multiplication laws:

\be G({\bf H}_{2,E}/\q)=G({\bf H}_{3,E}/\q)={\bf C}_2 \wr {\bf C}_3 \ee
\be (\gre_1 ,\gre _2 ,\gre _4; \tau_l) (\gre_1 ' ,\gre _2 ',\gre _4 '; \tau_{l '})= (\gre_{l'} \gre _ 1',\gre _{2l'}  \gre _ 2',\gre _{4l'} \gre _ 4'; \tau _{ll'}) \ee
and
\be \label{rgrg7} G({\bf H}_E/\q(\gr ))=\left({\bf C}_2 \times {\bf C}_2\right) \wr  {\bf C}_3 \ee

\be \label{mrgrg7} \left(\ba{ccc}
\gre_{2,1} ,& \gre _{2,2} ,&\gre _{2,4} \\
\gre_{3,1} ,& \gre _{3,2} ,&\gre _{3,4} \ea ;
 \tau _l \right) 
\left(\ba{ccc}
\gre_{2,1} ',& \gre _{2,2}' ,&\gre _{2,4}' \\
\gre_{3,1}' ,& \gre _{3,2}' ,&\gre _{3,4}' \ea ;
 \tau _{l'}\right)=
\ee 
\be \nonumber
= \left(\ba{ccc}
\gre_ {2,l'}\gre _ {2,1}' ,& \gre_{2,2l'} \gre _{2,2}',&\gre_{2,4l'} \gre _{2, 4}', \\
\gre_ {3,l'}\gre _ {3,1}' ,& \gre_{3,2l'} \gre _{3,2}',&\gre_{3, 4l'} \gre _{3,4}', \ea ;
 \tau_{ll'}\right). \ee

Remark. Observe, that instead of the triple of ideals ${\bf p}_3^k$ we could use in our investigations the 
second triple: ${\bf q}_3^k=(\tau ^{k-1}(5-3\gr))$ because, as we computed in Section \ref{arit}, 
the norms $N_{k/\q}(3+\gr)=13$ and $N_{K/\q}(5-3\gr)=83$ are primes different from $p=7$. Observe, that $13 \equiv 83 
\equiv -1 \,\, \text{mod}\,\, 7.$ Hence, we conclude that the primes $13$ and $83$ 
split completely in $\q(\gr) /\q$ whereas the prime ideals ${\bf p}_3^k$ and 
${\bf q}_3^k$ are inert for each $k$ in the field extension $\q(\go) / \q(\gr).$ 
On the other hand it is very well known, that $7$ is totally ramified in $\q(\go) / \q.$ 
Moreover, $N(2-\gr) = \prod_{k=1}^3 \tau ^{k-1} (2-\gr) = 7$, so the ideals 
${{\bf q}'}_3^k =(\tau^{k-1} (2-\gr))$ are all equal:
$$ (2-\gr)= (\tau  (2-\gr))=(\tau ^2 (2-\gr)).$$
Hence, we can not use the ideals by ${{\bf q}'}_3^k$ in our argument,
because
$ { {\bf q}'}_3 ^k | \GD _3 ^l$ for $ k,l=1,2,4$.


The Galois groups of compositions  of the extensions ${\bf H}_{\GL ,G}$ and $\q(\go) /\q$ are the following wreath product
\be G({\bf H}_{2,G}/\q)=G({\bf H}_{3,E}/\q)={\bf C}_2 \wr_{\phi} {\bf C}_6,\ee
\be \label{zgrg7} G({\bf H}_G/\q)=\left({\bf C}_2 \times {\bf C}_2\right) \wr_{\phi} {\bf C}_6, \ee
where $\phi (\pm k^2)=k^2 \mod 7$.
The multiplication laws in these groups are given by the following formulas
\be (\gre_1 ,\gre _2 ,\gre _4; \tau _l) (\gre_1 ' ,\gre _2 ',\gre _4 ';\tau _{l'})= (\gre_{\phi (l')} \gre _ 1',\gre _{2\phi(l') }  \gre _ 2',\gre _{4\phi(l')} \gre _ 4';\tau_{ ll'}) \ee
for $G({\bf H}_{r',G}/\q)$ and 
\be \label{mzgrg7}
\left(\ba{ccc}
\gre_{2,1} ,& \gre _{2,2} ,&\gre _{2,4} \\
\gre_{3,1} ,& \gre _{3,2} ,&\gre _{3,4} \ea ;
 \tau _l \right) 
\left(\ba{ccc}
\gre_{2,1} ',& \gre _{2,2}' ,&\gre _{2,4}' \\
\gre_{3,1}' ,& \gre _{3,2}' ,&\gre _{3,4}' \ea ;
 \tau _{l'}\right)=\ee
$$\left(\ba{ccc}
\gre_ {2,\phi (l')}\gre _ {2,1}' ,& \gre_{2,2\phi (l')} \gre _{2,2}',&\gre_{2, 4\phi (l')l} \gre _{2, 4}', \\
\gre_ {3,\phi (l')}\gre _ {3,1}' ,& \gre_{3,2\phi (l')} \gre _{3,2}',&\gre_{3, 4\phi (l')} \gre _{3, 4}', \ea ;
 \tau_{ll'}\right) $$
for $G({\bf H}_G/\q)$.

 
\section{Galois qubits and actions of the Galois group}
\label{galqu}

The matrix elements of projection operators in the Galois wavelets belong to the cyclotomic field $\q (\go)$. Also the wavelets are linear combinations of the arithmetic basis vectors over $\q (\go)$.
The matrix elements of density matrices in the wavelets belong to the  field $\q (\go, \sqrt{\GD _{r'}^k})$.  From columns of density matrices one gets the eigenbasis in the qubit $\ph _{r,r'}^k$.
This fact allows us to define the Galois-Heisenberg qubit as the space of linear combinations of these vectors over the field $\q(\go ,\sqrt{\GD _{r'}^k})$:
\be {\cal Q}_{r,r'}^k ={\cal Q}_{r,r'}^k( \q (\go ,\sqrt{\GD _{r'}^k}) ).\ee
The Galois-Heisenberg qubit $ {\cal Q}_{r,r'}^k$ is a two-dimensional bilinear space over the field ${\bf H}_{G,r'}^k:=\q (\go ,\sqrt{\GD _{r'}^k})$.
The density matrices can be also treated as endomorphisms of Galois qubits.
Let 
\be g=\left(\ba{ccc}
\gre_{2,1} ,& \gre _{2,2} ,&\gre _{2,4} \\
\gre_{3,1} ,& \gre _{3,2} ,&\gre _{3,4} \ea ;
 \tau _l \right) \in G({\bf H}_G/\q ).\ee
The  Galois group permutes the subfields
\be\label{permcial} g{\bf H}_{r',G}^k={\bf H}_{r',G}^{lk},\ee
and the eigenenergies
\be g E_{r',\nu}^k=E_{r',\gre _k \nu}^{lk}.\ee
Further, observe, that the Galois group commutes with the complex conjugation
\be \gk{ga}=g \gk{a}.\ee
Now let us define $V$ as the space over $\q$ with the arithmetic basis $|Q, {\bf j}\ket$, ${\bf j}\in Q_r, r=0,1,\cdots, 7$,
and operators of the space ${\bf H}_G \otimes _{\q } V$ and its endomorphisms in the following way:
\be \label{thetatheta} \theta _g =g \otimes _{\q} \id _V , \; \; \Theta _g =g\otimes _{\q} \id _{\mathrm{End}(V)}. \ee
Observe, that the operators defined in (\ref{thetatheta}) are linear over $\q$ but not over ${\bf H}_G $.

The scalar product is equivariant with respect to the Galois group action
\be \bra  \theta _g a, \theta _g b\ket=g\bra a, b \ket \ee
and  commutes with the hermitian conjugation of operators
\be (\Theta _g A)^{\dagger}=\Theta _g  A^{\dagger}.\ee
The Galois action on Galois wavelets satisfies the following formula
\be \theta _g |G,r,k,{\bf t}\ket=|G,r,lk, {\bf t}\ket. \ee
So the natural action of the Galois group permutes operators $S_{r,\GD r}^k$  in the way
\be \Theta _g  S_{r,\GD r}^k=S_{r,\GD r}^{lk}.\ee
Indeed, elements  of matrices of these operators in the wavelets belong to $\q (\go)$ and the matrices transform in the same way, while the action $\theta $ permutes the Galois qudits
\be  \theta _g{\cal Q}_{r,r'}^k={\cal Q}_{r,r'}^{lk}.\ee
Similarly, the action $\Theta $ permutes projection operators
\be \Theta _g  P_{r,r'}^k=P_{r,r'}^{lk}\ee
and density matrices
\be\label{permmg}  \Theta _g \vgr_{r,r',\nu}^k=\vgr_{r,r',\gre _k \nu}^{lk}.\ee


\section{Final remarks and conclusions}

We have presented here the solution of the eigenproblem of the Heisenberg Hamiltonian for the heptagonal chain in a precise, and algebraically exact form. To this aim, we have exploited some properties of finite extensions of the prime field $\mathbb{Q}$ of rationals.

The initial secular matrix for the Heisenberg Hamiltonian is purely arithmetic, i.e. its entries are integers. It reflects the fact that the set of positions of the system is finite, and encompasses $2^7=128$ elements. Introduction of quasimomenta requires, however, an extension of the field $\mathbb{Q}$ of rationals by the algebraic (but not arithmetic) integer $\omega=\exp{(2\pi i/7)}$, i.e. the cyclotomic field $\mathbb{Q}$. We have pointed out in a previous paper \cite{mbll} that in the case of pentagon $(N=5)$, the cyclotomic field $\mathbb{Q}(\omega)$ is already sufficient to express the whole eigenproblem. It is no more true for heptagon, since the characteristic polynomials of the secular matrix do not factorize fully in $\mathbb{Q}(\omega)$.

We have demonstrated here that one can control the classification of spectra and eigenstates of the system by use of arithmetic properties of finite field extensions. Moreover, these properties prove to be relatively simple on the case of heptagon since the resulting linear spaces are at most two-dimensional, and thus get an informatic interpretation of some qubits - memory units of a quantum computer, distinguished by Galois properties of the field extensions.


For the quasimomentum $k=0$ we gave directly the states and energies since the eigenproblem 
is very simple in this particular case.
To diagonalize the Hamiltonian for $k=0$ it is sufficient to use the field $\q$ ($r'=0,2,3$).

For quasimomenta $k\neq 0$ and $r'=1$ it is sufficient to use cyclotomic field $\q (\go)$ whereas for $r'=2,3$  
one needs to extend $\q(\go )$ by $\sqrt{\GD _{r'}^k}$.

For $r=1,2,3, \, r'=2,3, \, , r'<r$ we have constructed projectors $P_{r,r'}^k$ 
onto subspaces of states with given quantum numbers $r,r',k$. For $r'=2,3$ we have shown, that eigenenergies 
can not be expressed in the cyclotomic field $\q (\go)$.

We defined Heisenberg real fields ${\bf H}_{r',E}^k$, ${\bf H}_{r',E}$ and ${\bf H}_E$ 
as minimal fields that can be used to express eigenenergies for fixed $(r', k)$, $(r')$ and all eigenenergies
for the Heisenberg heptagon, respectively. Analogically, complex Heisenberg fields 
${\bf H}_{r',G}^k$, ${\bf H}_{r',G}$ and ${\bf H}_G$
are minimal ones that can be used to express eigenstates of the Hamiltonian and quasi-momenta
for fixed $(r', k)$, $(r')$ and for whole Heisenberg heptagon. 
Hence, in particular for $k=0$ all these fields are equal to $\q$.
For $ k\neq 0$ we have ${\bf H}_{1,E}^k=\q (\gr)$ and ${\bf H}_{r',G}^k=\q (\go)$ whereas
for $r'=2,3$ and $k\neq 0$ we have ${\bf H}_{r',G}^k=\q (\go, \sqrt{\GD _{r'}^k})$.
Moreover, we proved that $\sqrt{\GD_{r'} ^k} \notin \q(\gr)$ and, applying Kummer theory, we showed that
total real Heisenberg field ${\bf H}_E$ is an extension of $\q (\gr )$ 
by all six square roots $\sqrt{\GD_{r'}^k}$, $r'=2,3,$ $k=1,2,4$ and one cannot omit any of these 
roots to get the extension. The same holds for the Heisenberg complex fields but instead of the 
field $\q(\gr )$ we use $\q (\go )$.
Subgroups $G({\bf H}_E/\q (\gr ))= G({\bf H}_G/\q(\go ))=\z_2 ^6$ are generated by six reflexions
$\sqrt{\GD _{r'}^k} \na -\sqrt{\GD _{r'}^k}$ which fixed elements $\gr$ and $\go$. 
Galois groups $G({\bf H}_E/\q)$ i $G({\bf H}_G/\q)$ are presented as appropriate
wreath products (cf.  (\ref{rgrg7}, \ref{mrgrg7}) and (\ref{zgrg7}, \ref{mzgrg7}), respectively). 
We defined a notion of Galois and in particular Fourier-Galois, and Heisenberg-Galois qubit.
We introduced natural action of the Galois group $G({\bf H}_G/\q)$ and we showed how it permutes 
Heisenberg subfields ${\bf H}_{r',G}^k$, eigenenergies, projectors and density matrices. As it is clear from  
(\ref{permcial})-(\ref{permmg}) this action does not change quantum numbers $r,r'$ and relative positions vector 
${\bf t}$, however it permutes transitively the interior of the Brillouin zone and the 
digits $\nu=\pm 1$. The Galois group fixes $k=0$ and thus fixes the 
energies and density matrices for $k=0.$ It is important to observe, that this action restricted to the 
subgroup $G({\bf H}_G/\q(\go ))$ fixes wave vectors $k$, but does not fix
digits of qubits: $\nu \na \gre_l \nu$.\\

\noindent
{\bf Acknowledgements.} G.B. and J.M. were partially supported by the Polish NCN grants 
with numbers NN 201 607440 and 
NN 201 373236 respectively.


\end{document}